\documentclass{IEEEtran}
\usepackage{cite}
\usepackage{amsmath,amssymb,amsfonts}
\usepackage{algorithmic}
\usepackage{graphicx}
\usepackage{textcomp}
\usepackage{algorithm}
\def\BibTeX{{\rm B\kern-.05em{\sc i\kern-.025em b}\kern-.08em
    T\kern-.1667em\lower.7ex\hbox{E}\kern-.125emX}}

\usepackage{fancyhdr} 
\usepackage{soul}

\begin{document}
\title{Application of Iterative LQR on a Mobile Robot With Simple Dynamics}
\author{A. Aaqaoui (aaqaoui.ayo@gmail.com), MS Student \\  Y. M. Elsheikh M. (yoelsheikh@gmail.com), MS Student \\ Chair of Electromobility, TU Kaiserslautern}

\maketitle

\begin{abstract}
The aim in this paper is to apply the iLQR, iterative Linear Quadratic Regulator, to control the movement of a mobile robot following an already defined trajectory. This control strategy has proven its utility for nonlinear systems. As follows up, this work intends to concertize this statement and to evaluate the extent to which the performance is comparatively improved against the ordinary, non-iterative LQR. The method is applied to a differential robot with non-holonomic constraints. The mathematical equations, resulting description and the implementation of this method are explicitly explained, and the simulation studies are conducted in the Matlab and Simulink environment.    
\end{abstract}


\section{Introduction}
\label{sec:introduction}

\subsection{Linear Quadratic Regulator}
The Linear Quadratic Regulator (LQR) control is a modern state-space technique for designing optimal dynamic regulators. It refers to a linear system and a quadratic performance index according to
\begin{align}\dot{x}(t)&=Ax(t)+Bx(t).\label{eq} \\
x(0)&=x_{0}\end{align}
It enables a trade-off between the regulation performance and the control effort via the a performance index when the intial state \begin{equation}x_{0}\end{equation} is given. 
LQR assumes that the system is linear and hence expressed by
\begin{equation}\dot{x}=Ax+Bu.\label{eq}\end{equation}
Secondly, it assumes that the cost function, i.e. the performance index, is of the form
\begin{equation}J = \int_{0}^{\infty} [({x-x^{\ast}})^\intercal Q ({x-x^{\ast}}) + u^\intercal R u]dt.\label{eq}\end{equation}
where,
\begin{equation}Q=Q^\intercal \geq 0 \end{equation} and \begin{equation}R=R^\intercal \geq 0\end{equation}

$x^{\ast}$ is the target state; $Q$ and $R$, denote the state and input weighing matrices, respectively. \cite{b1} \cite{b4}

\subsection{Why iLQR}

Even in cases of non-linearity the LQR can provide optimal control, but its limitation is that the calculation of this optimal control law is conducted considering the local set of parameters without consideration of the generated changes in the future states. From this perspective the strength of the iterative linear quadratic regulator iLQR can be noticed. \par
The iLQR is an extension of the LQR control with idea of providing an optimal control input sequence considering the whole control sequence rather than only the control point at the current time-step. So this control strategy allows us to estimate an overall optimal sequence taking into account the changing dynamic of the system.  
The basics of this method can be described via the following algorithm: \cite{b2}
\begin{enumerate}
\item Initialization with: initial state $x_0$ and initial control input $U=[u_0,...,u_1]$
\item Forward Pass: applying the control sequence $U$ starting from $x_{0}$ to get the trajectory from the state space $x$.
\item Backward Pass: in this step the value function and the dynamics are evaluated for each $(x,u)$ of the state space.
\item Update control sequence calculation and $U^{'}$ and evaluating the trajectory cost from $(x_{0}, U^{'})$ then proceed a decision making of adopting U' based on $d=|cost(U)-cost(U^{'})$ and a threshold value|:
\begin{itemize}
    \item If $d < threshold$ ; convergence then we exit.
    \item If $cost(U) < cost(U^{'})$ ; we set $U = U^{'}$, and change the update size to be more aggressive, then go to step 2.
    \item If $cost(U) >= cost(U^{'})$ ; change the update size to be more modest, then go to step 3.
\end{itemize}

\end{enumerate} \par

 As for linear dynamics,in case of non-linearity is described analogously with the following equation and the non linearity is manifested via the function 
 $f$:
  \begin{equation}
    x_{k+1}=f(x_{k},u_{k})
  \label{eq}\end{equation}
 The equation describes the evolution of the vector state $x \in \mathbb{R}^n$ between times $i$ and $i+1$ given the control input $u\in \mathbb{R}^m$ at time i. And the resulting trajectory is denoted as \{$X$,$U$\}, where \{X\} is the sequence of state $X={x_{1},x_{2},..,x_{N}}$ and \{U\} is the corresponding control inputs $U={u_{1},u_{2},..,u_{N-1}}$. The underlying idea is that the generated input sequence is updated in every time-step for the to-go path. \cite{b2}

\subsection{Mobile Robot}
The aim in this paper is to develop a trajectory tracking controller that follows the a Linear Quadratic Regulator in essence and simulate its performance while concurrently, developing an iLQR for the same trajectory tracking problem. The two controllers are to then be compared as per the global trajectory tracking for a wheeled mobile robot. The chosen mobile robot is a differential robot with a castor set of wheels. 

The movement of the robot in the configuration space with respect to a generated global trajectory is denoted as nominal state sequence $S_{ff}={s_{1},s_{2},..,s_{N}}$ is controlled via its coordinates $(x,y)$ in the world frame and its orientation angle $\theta$ as illustrated in {fig1}.

\begin{figure}[!t]
\centerline{\includegraphics[width=\columnwidth]{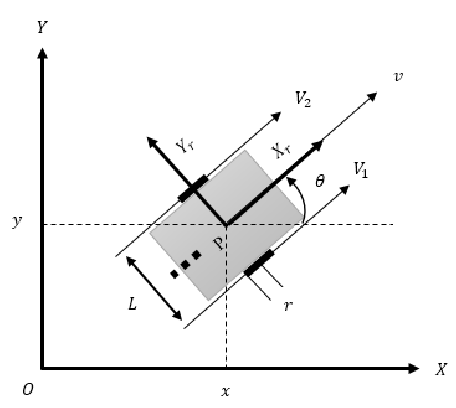}}
\caption{Differential Mobile Robot Parameters, courtesy of Yacine Ahmine's paper [6].}
\label{fig1}
\end{figure}

From $S_{ff}$ the nominal control sequence is derived by computing the needed linear velocity and angular velocity needed to reach state $s_{k}$ from the preceding state $s_{k-1}$ as shown in (x):
   
\begin{equation}
    u_{k} = \begin{bmatrix} v_{k}\\ \omega_{k} \end{bmatrix} =
    \begin{bmatrix} \frac{\sqrt{(x_{k}-x_{k-1})^{2}+(y_{k}-y_{k-1})^{2}}}{dt}\\ \frac{\omega_{k}-\omega_{k-1}}{dt} \end{bmatrix}
\end{equation}

The nominal control sequence is represented by the following vector:
$$ u_{ff} = {u_1,u_2,..,u_N} $$
where its components $u_k$ represent the nominal control from state $s_k$ to state $s_{k+1}$. \cite{b3}

\subsection{Approach}
The basic approach to a path tracking problem, considering a robot with non-holonomic constraints, is that the vehicle detects the environment through its sensors and then positions itself accordingly, setting the posture. After introducing the kinematics of the robot, the mathematical formulation is presented and the path tracking problem is then put forth in detail, along with the control strategy.
In accordance to the defined path, two control sequences are then produced and applied on the mobile robot.

\section{Mathematical Background}

\subsection{Cost Function and Value Function}
The total cost function of a nonlinear discrete system is written in quadratic form as:

\begin{equation}
\begin{aligned}
J_0 &= \frac{1}{2} (x_{N}-x^{*})^{T} Q_{f} (x_{N}-x^{*}) \\ 
& + \frac{1}{2} \sum_{k=0}^{N-1} ({x_{k}}^{T} Q x_{k} + {u_{k}}^{T} R u_{k} \end{aligned}
\end{equation}

It can be alternatively expressed as in equation (11), where $l$ represents the immediate cost at each state in the trajectory and $l_f$ the final cost:

\begin{equation}
    J(x_0,U) = \sum_{k=0}^{N-1} l(x_k,u_k) + l_{f}(x_{N}) ,
\end{equation}

the $x^{*}$ denotes the target state and $x^{N}$ the final state of the states sequence, $Q_f$ and $Q$ represent the state cost-weighting matrices which are symmetric positive semi-definite, and $R$ the control cost-weighting matrix which is positive definite. \par
In iLQR as indicated by its name the algorithm for finding the optimal trajectory is iterative. The aim is to compute an improved control input sequence by iterative simulation, from an initial state and a nominal control until convergence to an optimal trajectory. The cost function will indicate how the produced trajectory is deviated from the targeted trajectory by summing the resulting deviation at each states points. From linearization process, the locale deviation at state $x_k$ satisfy the following state equation: 
\\
\begin{equation}
    \delta x_{k+1} = A_{k} \delta x_{k} + B_{k} \delta u_{k} ,      
\end{equation}
\\
So the optimization formulation in iLQR can be written as follows:
\\
\begin{equation}
\begin {aligned}
     &\min_{u(t)} \frac{1}{2} (x_{N}-x^{*})^{T} Q_{f} (x_{N}-x^{*})\\
     & + \frac{1}{2} \sum_{k=0}{N-1} ({x_{k}}^{T} Q x_{k} + {u_{k}}^{T} R u_{k} ) 
     \end{aligned}
\end{equation} 
       such that:  $\delta x_{k+1} = A_{k} \delta x_{k} + B_{k} \delta u_{k}$ \\

In order to ease the evaluation of "the minimum", we define, from a specific state $x_t$ (or at time $t$) in the trajectory, the cost-to-go as the sum of costs from time $t$ to the last state $x_N$ with the corresponding input sequence $U_t={u_{t},u_{t+1},..,u_{N-1}}$:

\begin{equation}
   J_t(x,U_t) = \sum_{k=t}^{N-1} l(x_k,u_k) + l_{f}(x_{N}) , 
\end{equation}
 
The optimal cost-to-go from this given state is denoted as the value function $V$ at time $t$, and Thus expressed as follows:

\begin{equation}
\begin{aligned}
    V_{t}(x) &= \min_{U_t} J_t(x,U_t)  ; 
    & t < N \\
    V(x_N) &=l_{f}(x_{N}) ;
    & t=N 
    \end{aligned}
\end{equation}

as the $V(x_N)$ can be determined independently of the control input, comes the idea to evaluate the value function in a sequence starting backward from target state and proceeding recursively to the preceding state and so on. This second formulation of the value function shows this possibility, and it is written as a function of the immediate cost $l(x,u)$ and the value function in the next time-step:

\begin{equation}
    V_{t}(x) = \min_{u} [l(x,u) + V(f(x,u))]  
\end{equation} \cite{b5}

 then the argument of the minimum in last equation is introduced in terms of perturbations in order to find, at each pair $(x,u)$, the supplement control input $\delta u$ that would minimize the perturbations expressed as follows:
 
 \begin{equation}
 \begin{aligned}
    Q(\delta x,\delta u) &= l(x+\delta x,u+\delta u)\\ &- l(x,u) + V(f(x+\delta x,u+\delta u))\\ &- V(f(x,u)) , 
    \end{aligned}
\end{equation}
via Taylor's expansion it can be approximated under this form 
\begin{equation}
    Q(\delta x,\delta u) =\frac{1}{2} \begin{bmatrix}\\ \delta x \\ \delta u \end{bmatrix}^T \begin{bmatrix} 0& Q_{x}^T& Q{u}^T\\ Q_{x}& Q{xx}& Q{xu}\\ Q_{u}& Q{ux}& Q{uu} \end{bmatrix} \begin{bmatrix} 1\\ \delta x \\ \delta u \end{bmatrix}
\end{equation}
\cite{b5}

where the expansion coefficients are :
\begin{align*}
Q_{x} &=l_{x}+f_{x}^T V_{x}^{'}\\
Q_{u} &=l_{u}+f_{u}^T V_{x}^{'}\\
Q_{xx} &=l_{xx}+f_{x}^T V_{xx}^{'}f_{x}+V_{x}^{'}f_{xx}\\
Q_{uu} &=l_{uu}+f_{u}^T V_{xx}^{'}f_{u}+V_{x}^{'}f_{uu}\\
Q_{ux} &=l_{ux}+f_{u}^T V_{xx}^{'}f_{x}+V_{x}^{'}f_{ux}
\end{align*}

the control updating term is computed as follow :
\begin{equation}
    \delta u^{*} = \min_{\delta u} Q(\delta x,\delta u) = -Q_{uu}^{-1}(Q_{u}+Q_{ux}\delta x), 
\end{equation}

which can be divided into two terms; an open-loop term $k=-Q_{uu}^{-1}Q_{u}$ and a feedback gain term $K=-Q_{uu}^{-1}Q_{ux}$.Plugging the policy into the expansion of perturbations we can access to a quadratic model of the value at time $i$:
\begin{align}
\Delta V(i)&=-\frac{1}{2} Q_{u}Q_{uu}^{-1}Q_{u}\\
V_{x}(i)&= Q_{x}- Q_{u}Q_{uu}^{-1}Q_{ux}\\
V_{xx}(i)&= Q_{xx}- Q_{xu}Q_{uu}^{-1}Q_{ux}
\end{align}

now we are able to start the process again from time step $i-1$. a recursive computing work is then conducted to derive at each point the local quadratic models of $V(i)$ and the control modifications ${k(i),K(i)}$ which represents the output of the backward pass.After completion a new forward pass computes a new trajectory with new updated control sequence. \cite{b5}

\begin{align}
\hat x(1)&=x(1)\\
\hat u(i)&=u(i)+k(i)+K(i)(\hat x(i)-x()\\
\hat x(i+1)&= f(\hat x(i),\hat u(i))
\end{align}

\begin{figure}[ht]
\centerline{\includegraphics[width=\columnwidth]{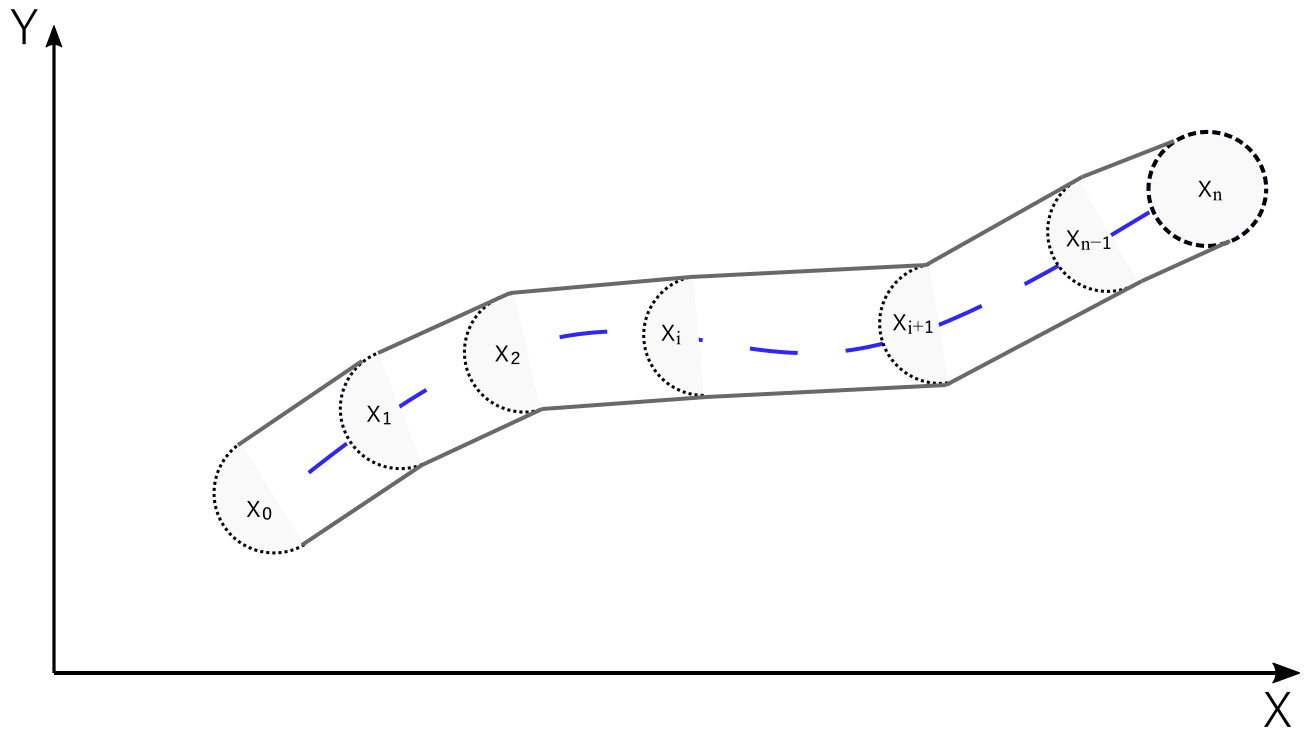}}
\caption{Schematisation of the iLQR Working Principle.}
\label{fig2}
\end{figure}

\subsection{Algorithms}

First with LQR Approach:

\begin{figure}[ht]
\centerline{\includegraphics[width=\columnwidth]{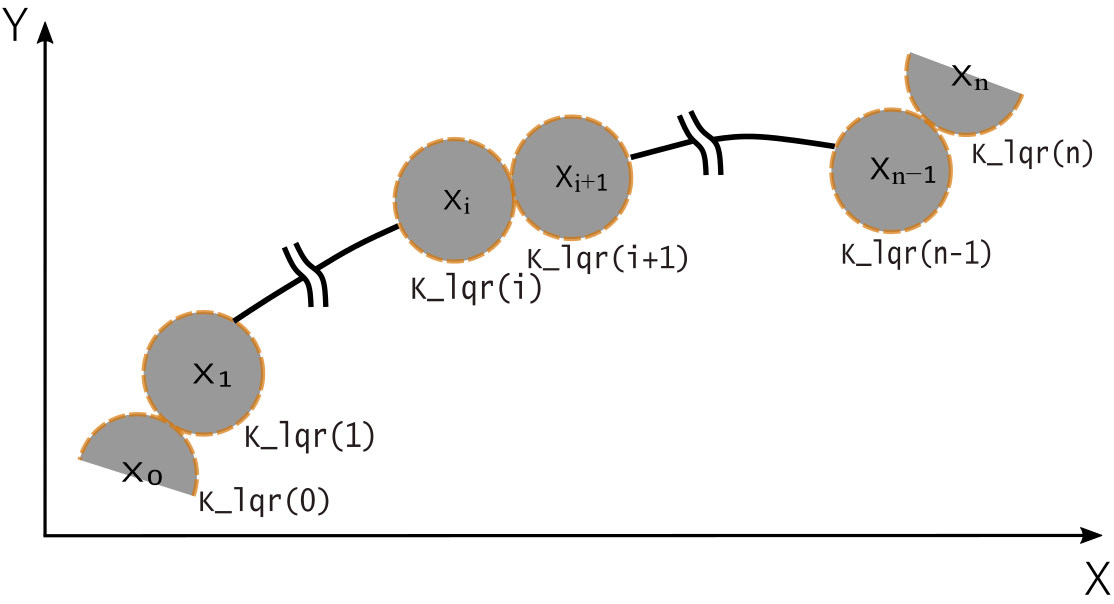}}
\caption{Schematizing of the LQR approach.}
\label{fig3}
\end{figure}

\begin{algorithmic}[1]
\STATE set initial state and target state, $x_0$ and $x_N$.
\STATE set remaining states and compute corresponding control sequence: $(X^*,U^*)=([x_1,..,x_N] , [u_1^*,..,u_{N-1}^*])$.
\\ *Linearization \& Updating control sequence:
\FOR{$i=1; i<N, i++$}
\STATE Compute Jacobiens $J_x$, $J_u$ at operating points $S_i=(x_i,u_i)$.
\STATE LQR formulation around $S_i$:$K_i = lqr(A_i,B_i,C_i,D_i,Q,R)$, 
\\ with  $A_i=J_x$ ; $B_i=J_u$ ; $C_i=I$ ; $D_i=0$ 
\STATE compute control law :  $\delta u = K_i\delta x$
\STATE updating control input at $S_i$ :  $u_(i) = u_i^* + \delta u_i$
\ENDFOR

\end{algorithmic}
\par
Then with iLQR Approach :

\begin{algorithmic}[1]
 \STATE evaluate initial input $V_{N}$ and then $V_{N,x}$, $V_{N,xx}$\\
 \STATE recall the computed values in forward pass of derivatives of local cost dynamic function at time step $i$ : ${l_{x}(i),l_{u}(i),l_{xu}(i),..}$ and ${f_{x}(i),f_{u}(i),f_{xu}(i),..}$\\
 \STATE deduction of different derivatives of value perturbation at time step $i$ ${Q_{x}(i),Q_{u}(i),Q_{xu}(i),..}$ \\
\STATE compute control input update; feed-forward term $k$ and feedback term $K$ : ${k{i},k{i}}$ \\
\STATE evaluate $V_{i,x}$, $V_{i,xx}$ \\
\STATE set i - - \\
\end{algorithmic}

\section{Example of application}

\subsection{Kinematic Model}
the movement of the differential mobile robot can be described with simple kinematic model emanating from its principle of manoeuvring. Its movement in the configuration space is ensured only by the rotation of the wheels around their axis (no steering). To produce a linear movement the wheels have to rotate with same velocity and when they have different angular velocities the robot's body will rotate correspondingly to this difference. to quantify it's movement in any unknown environment (planar), we need the values of three parameters:
$(x,y)$ indicating the coordinate of the robot center of mass, and $\theta$ the orientation which the angle between the line perpendicular to the wheel axis and the X-axis as shown in the fig (1)
these parameters are satisfying the following equations and thus providing the kinematic model of the differential robot.
\\  
\begin{equation}
\begin{aligned}
    \dot{x} &= v cos{\theta} \\
    \dot{y} &=v sin{\theta} \\
    \dot{\theta}&=\omega
\end{aligned}
\end{equation}
\\
For tracking, as shown in the equation, the robot is controlled by the input u, which is composed by the linear velocity $v$ and angular velocity $\omega$.
\\ $$ u = [v , w]^{T}$$

since we are dealing with a nonlinear system, the sensors information about the position will captured as they are required in the linearization process.As a result, the state vector $x$ is including extra necessary states: $ u = [x,y,\theta,v,\theta_{k+1}\\,\omega,\delta \omega]^{T}$. \cite{b3}
\begin{equation} 
    x_{k+1} = \begin{bmatrix} x_{k+1}\\ y_{k+1}\\ \theta_{k+1}\\ v_{k+1}\\ \omega_{k+1}\\ \delta v_{k+1}\\ \delta \omega_{k+1}\end{bmatrix} =
    \begin{bmatrix} x_{k} + v_{k}cos\theta dt \\ y_{k} + v_{k}sin\theta dt \\ \theta_{k} + \omega_{k} dt \\ u_{k}(0) \\ u_{k}(1) \\ u_{k}(0) - v_{k} \\ u_{k}(0) - \omega_{k} \end{bmatrix} =
    f(x_{k},u_{k}) 
\end{equation}

\section{Simulation and discussion}
In matlab-simulink environment a path in form of inclined bell shape is constructed to conduct the simulation. it is a general very encountered form in real application, and a form with which steering performance can be checked. the two case scenarios are simulated: in the first scenario the ordinary LQR strategy is deployed for following the defined path, while in the second scenario the iLQR approach is considered. the following figures show the resulting path from both cases.\par
Both  scenarios provide good tracking behavior with optimal error, with slight improvement with the case of iLQR. The resulting path using iLQR method is not fully reflecting its advantage because the optimization calculation are based on the discrete points represented by the operating points, whereas the calculation conducted in the LQR case is not quite similar since the controller is updating the control input within time between two operating points. A proper comparison would be by considering same rate of updating control, and this it would be taken into account in future work.  Furthermore, in simulation we notice the impact of the initial control input quality and also the impact of the number of points constituting the path; the higher their number the more the tracking is stable and approaching the targeted path.\par

\begin{figure}[h]
\centerline{\includegraphics[width=\columnwidth]{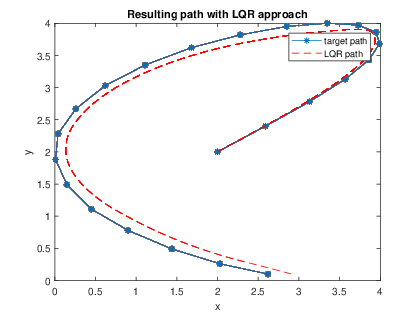}}
\caption{Resulting Path with LQR approach.}
\label{fig5}
\end{figure}

\begin{figure}[h]
\centerline{\includegraphics[width=\columnwidth]{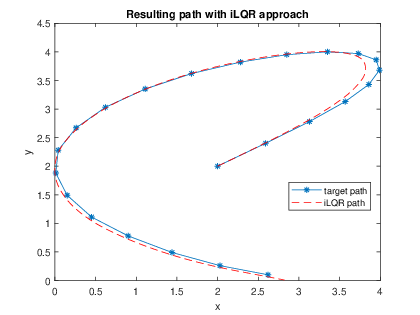}}
\caption{Resulting Path with iLQR approach.}
\label{fig5}
\end{figure}

\subsection{Conclusion}
In this work we present the theoretical aspect behind the iLQR approach for trajectory tracking control in case of non linear dynamics and see how its concept  encompasses the concepts of LQR approach by considering the total cost. Works as [3] and [5] had shown experimentally the powerfulness of this approach and provide considerable insights for this work. In this paper a simulation was conducted by taking the dynamic of a differential mobile robot as an example of application. The results  show that he approach gives  stable solution with minimal errors.


\end{document}